\documentclass[espf]{aa}

\usepackage{color} 

\usepackage{graphicx}
\usepackage{amsmath}
\usepackage[psamsfonts]{euscript}
\usepackage[psamsfonts]{amssymb}
\usepackage{amsmath}
\usepackage{graphicx}
\usepackage{natbib}
\bibpunct{(}{)}{;}{a}{}{,} 

\usepackage{epsfig}
\titlerunning{}

\def\kms{km~s$^{-1}$}

\begin{document}
\title{Understanding the dynamical structure of pulsating stars:\\ The Baade-Wesselink projection factor of the $\delta\,$Scuti stars \\ AI Vel and $\beta\,$Cas \thanks{This work uses observations made with the HARPS instrument at the 3.6 m telescope (La Silla, Chile) in the framework of the LP185.D-0056 and with the SOPHIE instrument at OHP (France).}}

\titlerunning{The Baade-Wesselink projection factor of the $\delta\,$Scuti stars}
\authorrunning{G. Guiglion and collaborators}

\author{G. Guiglion \inst{1}, N. Nardetto \inst{1}, P. Mathias \inst{2}, A. Domiciano de Souza \inst{1}, E. Poretti \inst{3}, M. Rainer \inst{3}, A. Fokin \inst{4}, D. Mourard \inst{1}, W. Gieren \inst{5}}

\institute{Laboratoire Lagrange, UMR 7293, UNS/CNRS/OCA BP 4229, 06304 Nice Cedex 4, France \and Institut de Recherche en Astrophysique et Plan\'etologie, UMR 5277, 57 avenue d'Azereix, 65000 Tarbes, France \and INAF-Osservatorio Astronomico di Brera, Via E. Bianchi 46, 23807 Merate, Italy \and Institute of Astronomy of the Russian Academy of Sciences, 48 Pjatnitskaya Str., Moscow 109017, Russia \and Departamento de Astronom\'ia, Universidad de Concepci\'on, Casilla 160-C, CL Concepci\'on, Chile}

\date{Received 23 November 2012; accepted 2 January 2013}

\abstract{} {The Baade-Wesselink method of distance determination is based on the oscillations of pulsating stars. The key parameter of this method is the projection factor used to convert the radial velocity into the pulsation velocity. Our analysis was aimed at deriving for the first time the projection factor of $\delta\,$Scuti stars, using high-resolution spectra of the high-amplitude pulsator AI Vel and of the fast rotator $\beta\,$Cas.} {The geometric component of the projection factor (\emph{i.e.} $p_{\text{0}}$) was calculated using a limb-darkening model of the intensity distribution for AI Vel, and a fast-rotator model for $\beta\,$Cas. Then, using SOPHIE/OHP data for $\beta\,$Cas and HARPS/ESO data for AI Vel, we compared the radial velocity curves of several spectral lines forming at different levels in the atmosphere and derived the velocity gradient associated to the spectral-line-forming regions in the atmosphere of the star. This velocity gradient was used to derive a dynamical projection factor $p$.} {We find a flat velocity gradient for both stars and finally $p = p_0 = 1.44$ for AI Vel and $p = p_0 = 1.41$ for $\beta\,$Cas. By comparing Cepheids and $\delta\,$Scuti stars, these results bring valuable insights into the dynamical structure of pulsating star atmospheres. They suggest that the period-projection factor relation derived for Cepheids is also applicable to $\delta\,$Scuti stars pulsating in a dominant \textit{radial} mode.}{}


\maketitle

\section{Introduction}
Determining distances in the Universe is not a trivial task. From our Galaxy to the Virgo Cluster, distances can be derived using the period-luminosity relation ($PL$) of Cepheids \citep{riess_2009b, riess_2009a}. However, this relation has to be calibrated, using the Baade-Wesselink method of distance determination for instance \citep{storm_2011_a, storm_2011_b}. The principle of this method is simple: after determining the angular diameter and the linear radius variations of the star, the distance is derived by a simple ratio. Angular diameter variations can be measured using interferometry \citep{kervella_2004_inter} or the infrared surface brightness relation \citep{gieren_1998,gieren_2005}. The linear radius variation is measured by integrating the pulsation velocity (hereafter $V_{\text{puls}}$) over one pulsating cycle. However, from observations we have only access to the radial velocity ($V_{\text{rad}}$) because of the projection along the line-of-sight. The projection factor, used to 
convert the radial velocity into the pulsation velocity, is defined by $p = V_{\text{puls}} / V_{\text{rad}}$. There are in principle three sub-concepts involved in the Baade-Wesselink projection factor: (1)~the geometric projection factor $p_\mathrm{0}$, which is directly related to the limb-darkening of the star (see Sect.~\ref{geom_proj_fact}), (2)~the correction $f_\mathrm{grad}$ due to the velocity gradient between the spectral-line-forming region and the photosphere of the star; this quantity can be derived directly from observations by comparing different lines forming at different levels in the atmosphere (see Sect.~2 and~\ref{atmosph_corre}), and (3)~the correction $f_\mathrm{o-g}$ due to the relative motion between the {\it optical} and {\it gas} layers associated to the photosphere (see Sect.~\ref{atmosph_corre}). For a detailed analysis of the p-factor decomposition we refer to \cite{fog_nardetto_2007}. The projection factor is then defined by $p = p_\mathrm{0} f_\mathrm{grad} f_\mathrm{o-g}$. In the 
following, we apply this decomposition of the projection factor (originally developed for Cepheids) to the $\delta$ Scuti stars AI Vel and $\beta\,$Cas. 
The impact of non-radial modes of $\delta$ Scuti stars on the projection factor is a very difficult question studied previously \citep{Dziembowski_1977, Balona_1979, Stamford_1981, Hatzes_1996}. The particular cases of AI Vel and $\beta\,$Cas are discussed in the conclusion.
This paper is part of the international {\it Araucaria Project}, whose purpose is to provide an improved local calibration of the extragalactic distance scales out to distances of a few megaparsecs~\citep{gieren_2005}. In this context, $\delta$ Scuti stars are extremely interesting since it has been shown recently that they follow a  $PL$ relation \citep{Mcnamara_2007_delta_scuti, poretti_2008}.

\section{Spectroscopic observations of $\delta$ Scuti stars}\label{targets}

\begin{figure}[h!]
\centering
\includegraphics[width=0.8\linewidth]{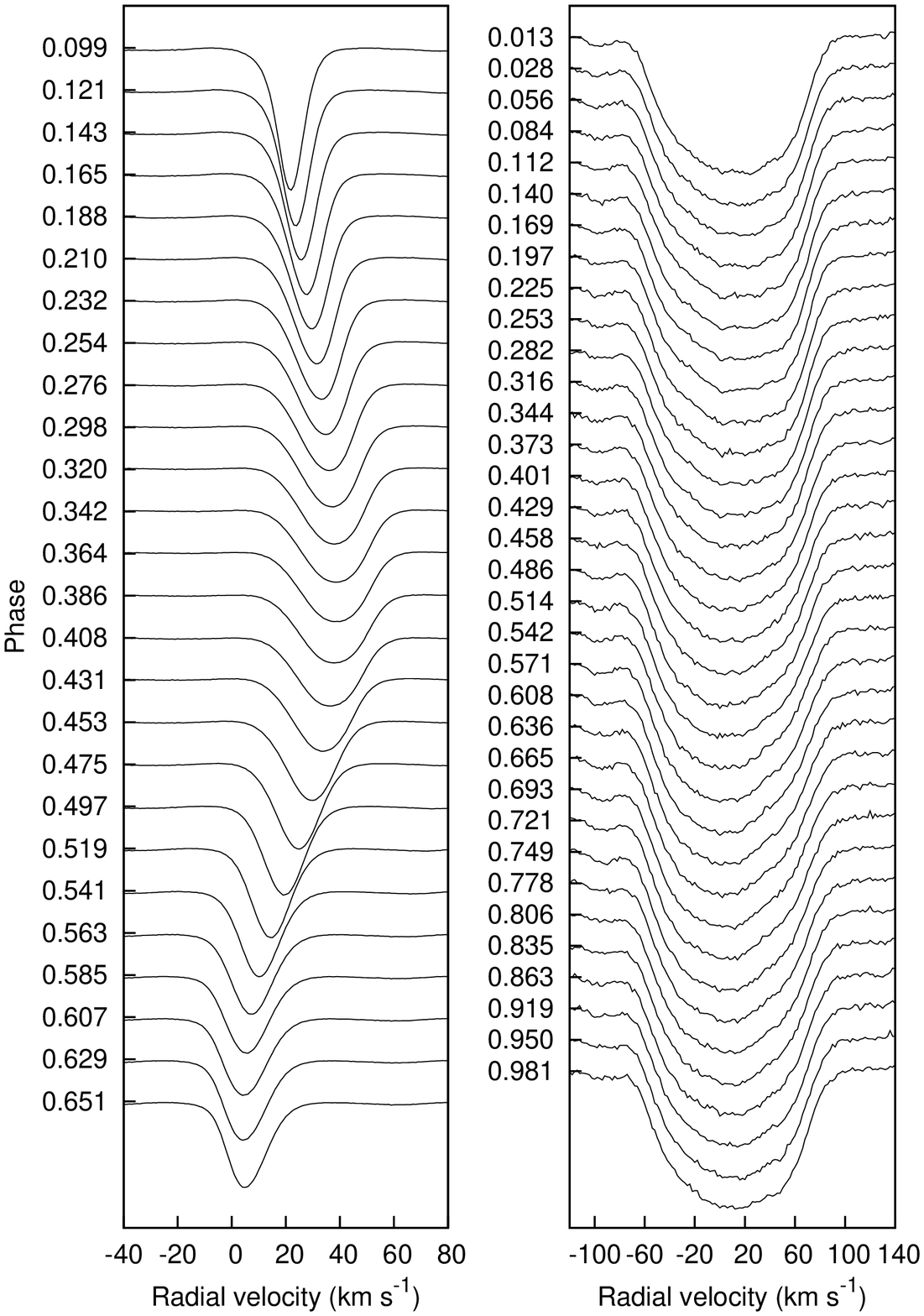}
\vspace{-0.5cm}
\caption{\label{film_aivel}Mean profiles of the HARPS spectra of AI Vel (\emph{left panel}, T$_0=$JD\,2443176.00) and of the SOPHIE spectra of  $\beta\,$Cas (\emph{right panel}, T$_0=$JD 2438911.88). The pulsation phase is given on the y-axis. A strong broadening is clearly seen for $\beta\,$Cas due to its high rotation velocity. We computed the mean line profiles of AI Vel and $\beta\,$Cas spectra by means of a deconvolution process using the LSD software \citep{donati_1997}. }
\end{figure}

AI Vel (HD\,69213, A9\,IV/V) is one of the most often observed double-mode, high-amplitude $\delta$\,Scuti stars. This star pulsates in the fundamental and first overtone radial modes with a well-constrained period ratio $P_1/P_0$ of 0.77 \citep{Poretti_2005}. In addition to $P_0=0.111574$\,d and $P_1=0.0862073$\,d, \citet{ephemeride_ai_vel} clearly detected two other periods, tentatively identified as the third and fifth radial overtones.
We observed AI Vel using the $\text{HARPS}$ spectrograph mounted at the ESO 3.6-m telescope. We  analysed  26 high signal-to-noise ratio (S/N$\simeq$140) spectra taken in the high-efficiency mode (EGGS, \text{R}=80\,000) on the night of January 9-10, 2011. We identified 53 metallic unblended spectral lines (ranging from $3\,780$ to $6\,910\,$\AA) relevant for the determination of radial velocities. Figure~{\ref{film_aivel}} (\emph{left panel}) shows the behaviour of the mean spectral line profile along the pulsation phase. The shifts due to the radial modes clearly dominate the line profile variations.
\vspace{-1.0cm}

The target $\beta\,$ Cas (HD\,432, F2\,III/IV) is a low-amplitude $\delta\,$Scuti star. We observed $\beta\,$Cas with the SOPHIE instrument ($\text{R}= 75\,000$) at the OHP 1.93-m telescope on the night of September 30, 2011. We collected 241 high-resolution spectra with a mean S/N of 100. We could distinguish only height unblended spectral lines relevant for the spectral analysis because of the strong rotational broadening (\figurename{~\ref{film_aivel}}, \emph{right panel}). \citet{riboni_1994_P} showed that the star is a monoperiodic pulsator at the detection limit of ground-based photometric measurements, with a pulsation period of $P=0.101036676\,$d. The mode identification  is unclear \citep{rodriguez_1992, riboni_1994_P}. Today, the distance of $\beta\,$ Cas is known to be 16.8~pc from the Hipparcos parallax \citep{van_Leeuwen_2007}. Thus we could obtain the absolute magnitude $M_V$=1.14 from  the apparent magnitude $V$=2.27. The $PL$ relations \citep{Mcnamara_2007_delta_scuti, poretti_2008} supply a fundamental radial period of about 0.15~d at this $M_V$ value. Therefore, the observed period is similar to that expected for the second radial overtone.   We attempted a mode identification from our spectroscopic data using the FAMIAS\footnote{Developed in the framework of the FP6 European Coordination Action HELAS (http://www.helas-eu.org/).} software. Since we dealt with a fast rotator, we used the Fourier parameter fit method \citep{Zima_2006}. Imposing the frequency $1/P$, the results from spectroscopy point towards an axisymmetric mode, without a clear indication on the $\ell$-value. Since $\beta\,$ Cas  is seen almost pole-on (\mbox{$i = 19.9 \pm 1.9 ^{\circ}$}, \citet{che_bet_cas_image_2011}), a low--degree, axisymmetric, nonradial mode mimics the pulsation behaviour of a radial mode very well. On the basis of these considerations, we treated  $\beta\,$ Cas as a radial monoperiodic pulsator for our purposes. We also used the mean line profiles of $\beta\,$ Cas to estimate the $v \sin i$ values from the position of the first zero of their Fourier transforms \citep{carroll_1933}. This approach is possible only for objects where the rotational broadening is dominant with respect to the other sources of line broadening (e.g., instrumental effect, microturbulence), which is always the case for $\beta\,$ Cas, but not for AI Vel, where we were unable to use the Fourier transform method on the narrower lines ($v \sin i < 10$ \kms). The radial velocity values of the observed profiles of $\beta\,$Cas range from 5.3 to 11.6~\kms\, and the $v \sin i$ values from 74.0 to 77.5~\kms\, (\figurename{\ref{film_aivel}}, \emph{right panel}). We could also determine a mean  $v\sin i$ value from the average of all the mean profiles and obtained $75.72 \pm 0.14$ \kms.  This value is consistent with literature values \citep{vsini_beta_cas_85kms,vsini_beta_cas_72kms,vsini_beta_cas_71kms}.
When considering \mbox{$i = 19.9^{\circ}$}, this means that $\beta\,$Cas is an intrinsic fast rotator, with a velociy of $v_{\text{rot}}\simeq\,$220~\kms~which is consistent with \citet{che_bet_cas_image_2011}.

Finally, for both stars, the centroid radial velocity $RV_{\text{c}}$ (or the first-moment radial velocity) and the line depth $D$ are derived as a function of the pulsation phase for each selected spectral line. These data are used in Sect. 4.

\section{The geometric projection factor $p_0$}\label{geom_proj_fact}

\begin{figure}[ht!]
 \centering
 \includegraphics[width=0.38\textwidth,clip]{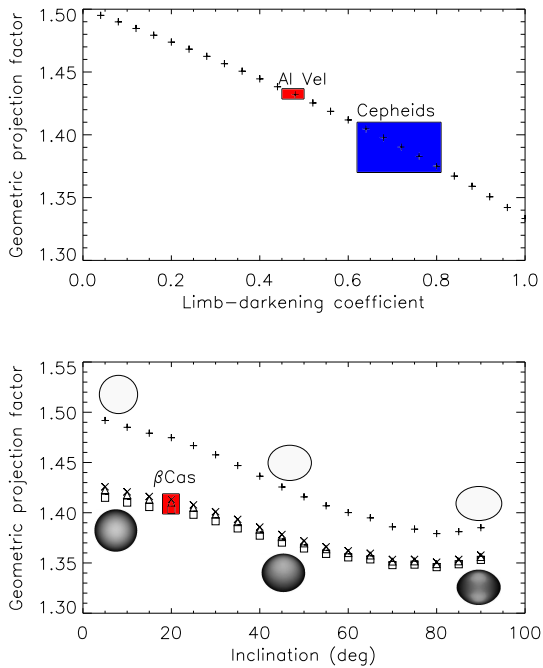}
 \caption{\emph{Top}: $p_\mathrm{0}$ as a function of the limb-darkening parameter $u_{\lambda}$. The red box indicates the uncertainty on $p_0$ for the $\delta$ Scuti AI Vel. The blue box indicates the typical values of $u_{\lambda}$ and $p_0$ for Cepheids. The dots corresponds to the relation provided by \cite{nardetto_2006}. \emph{Bottom}: \mbox{$p_\mathrm{0}$} as a function of the inclination of the fast rotating star $\beta\,$Cas for three different wavelengths ($\lambda=6\,000$~\AA~($\square$), $\lambda=6\,500$~\AA~($\bigtriangleup$), and $\lambda=7\,000$~\AA~($\times$)). The red box indicates the uncertainty on $p_0$ for $\beta\,$ Cas. The case of an uniform elonged disc is over-plotted ($+$), and we find that $p_0 = 1.5$ for $i = 0^{\circ}$, as expected for a circular uniform disc.}
 \label{u_p0_i}
\end{figure}

Considering a limb-darkened pulsating star in rotation with a one-layer atmosphere, the projection factor is purely geometric. Thus, $p = \frac{V_{\text{puls}}}{V_{\text{rad}}} = p_0$. The radial velocity is then defined by

\begin{equation}\label{eqvrad}
V_{\text{rad}}=\frac{1}{\pi \text{R}^2} \int_{x,y\,\in \text{D}_{\text{R}}} I(x,y,\lambda)\cdot V_{\text{puls}} \cdot \sqrt{1-\frac{\left(x^2+y^2\right)}{\text{R}^2}}\,\text{d}x\text{d}y,
\end{equation}

\noindent where $\text{D}_{\text{R}}$ is the surface of the stellar disc of radius R,  and $I(x,y,\lambda)$ the limb-darkened \textit{continuum} intensity distribution considered at the wavelength of observation $\lambda$ defined by $I(x,y,\lambda) = I_0 (1 - u_{\lambda} + u_{\lambda} \sqrt{1 -\left(x^2+y^2\right)})$, where $u_{\lambda}$ is the linear limb-darkening coefficient from \citet{claret_2011}. Considering $\text{T}_{\text{eff}}\,=\,7\,400\, \text{K}$ and $\text{log} \, g\,=\,3.5$, we find $u_{\text{R}} = 0.474 \pm 0.025$ in the $R$-band from \citet{claret_2011}. Using Eq. 1, we deduce a value of the geometric projection factor for AI Vel of $p_0 = 1.43 \pm 0.01$. $p_0$ is assumed to be constant with the pulsation phase \citep{nardetto_2004}. This value is higher than what we generally obtain for Cepheids (typically 1.37 to 1.41, see \figurename{\ref{u_p0_i}}~(\emph{top panel})).

The geometrical shape of $\beta\,$ Cas and its intensity distribution are distorted by its high rotation rate. The geometric projection factor depends on the inclination of the star's rotation axis compared to the line-of-sight. If the star's rotation axis is along the line-of-sight ($i=0^{\circ}$), the star is observed pole-on and is seen as a disc. For $i > 0^{\circ}$ the star has an ellipsoidal shape.

\begin{figure}[h!]
\centering
\includegraphics[width=0.8\linewidth]{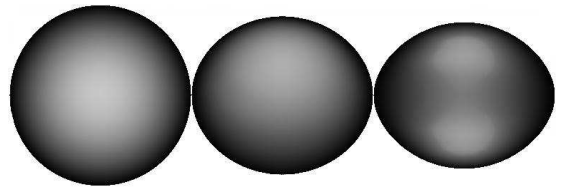}
\caption{\label{beta_cas_i}Modelled intensity distributions of $\beta\,$Cas at $6\,000\,$\AA~for $i=5^{\circ}$ (\emph{left}), $i=50^{\circ}$ (\emph{middle}), $i=90^{\circ}$ (\emph{right}).}
\end{figure}

Using the fundamental parameters of the modified von Ziepel model found by \citet{che_bet_cas_image_2011} and the rotating stars model by \cite{domiciano_2002_imap, domiciano_2012}, we derived the intensity distribution in the continuum for different inclinations of the star (from $i=0^{\circ}$ to $i=90^{\circ}$ with a step of $5^{\circ}$) and for three wavelengths: $\lambda = 6\,000, 6\,500 ~ \text{and} ~ 7\,000$~\AA. Using these intensity maps, we can easily calculate the geometric projection factor. Indeed, for an ellipsoid, $V_{\text{rad}}$ is then defined by

\begin{equation}\label{eqvrad_beta}
V_{\text{rad}}=\frac{1}{\pi \text{R}^2} \int_{x,y\,\in \text{D}_{\text{R}}} I(x,y,\lambda) \cdot V_{\text{puls}} \cdot \sqrt{1-\left(\frac{x^2}{\text{a}^2}+\frac{y^2}{\text{b}^2}\right)} \text{d}x\text{d}y,
\end{equation}

\noindent where a and b are the semi-major and semi-minor axis of the ellipse. In \figurename{~\ref{u_p0_i}}~(\emph{bottom panel}), we show the geometric projection factor ($p_0$) as a function of~$i$. \figurename{~\ref{beta_cas_i}} presents the modelled intensity distributions for several inclinations. This relation is extremely interesting because it shows that the inclination of a fast-rotating star can have an impact of more than 10\% on the projection factor. Of course, it also depends on the rotation velocity of the star: the higher the rotation velocity (for a given inclination), the lower the projection factor. Using the inclination found by \citet{che_bet_cas_image_2011}, \mbox{$i = 19.9 \pm 1.9 ^{\circ}$}, we finally find a geometric projection factor for $\beta\,$Cas of  $p_{\text{0}} = 1.41 \pm 0.02$ (averaged over the three wavelengths considered).\\
\vspace{-0.5cm}

\vspace{-0.5cm}

\section{Dynamical structure of the atmosphere}\label{atmosph_corre}

\begin{figure}[ht!]
 \centering
 \includegraphics[width=0.45\textwidth,clip]{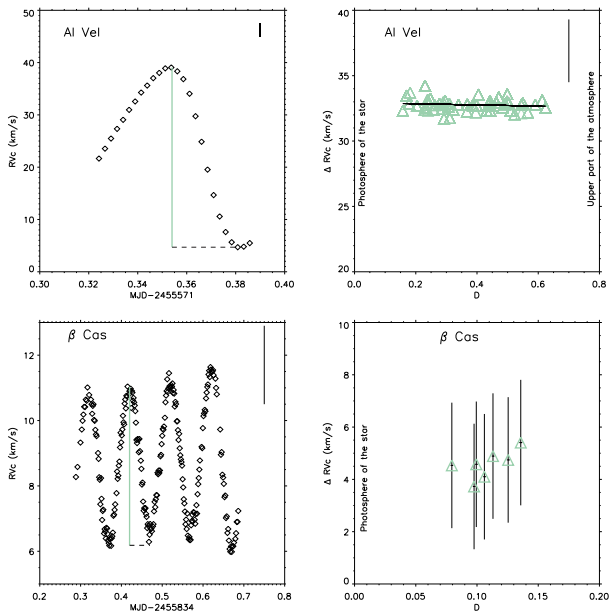}
 \caption{\emph{Top}: \mbox{$RV_\text{c}$} as a function of the MJD in the case of the FeII $5\,234.625\,$\AA~spectral line, and amplitude of the $RV_{\text{c}}$ curves as a function of the spectral line depth ($D$) for AI Vel. Typical error bars are indicated in the upper right of each panel. \emph{Bottom}: The same for $\beta\,$Cas f the FeII $4\,508.288\,$\AA~spectral line.}
 \label{triple}
\end{figure}

By comparing the 2K-amplitude (defined as the amplitude of the $RV_{\mathrm{c}}$ curve, hereafter $ \Delta RV_{\mathrm{c}}$) with the depth (D) of the 53 spectral lines selected in the case of AI Vel, one can directly measure the atmospheric velocity gradient in the part of the atmosphere where the spectral lines are formed  (Figure~\ref{triple}, top-left). To quantify the impact of velocity gradient on the projection ($f_{\text{grad}}$), we do not need to derive the velocity gradient over the whole atmophere, but only at the location of the forming regions of the spectral lines used to derive the distance of the star.  We therefore performed a linear regression  according to the relation \mbox{$\Delta RV_{\text{c}} = a_0\,D+b_0$}. We obtained $\Delta RV_{\text{c}} =  [-0.40 \pm 0.53]D+[32.87 \pm 0.23]\,\text{km}\,\text{s}^{-1}$ (Fig.~\ref{triple}, top-right). In principle, $f_{\text{grad}}$ depends on the spectral line considered \citep{fog_nardetto_2007}: $f_{\text{grad}} = b_0 / (a_0D+b_0)$. Here, we find that $f_{\text{grad}}$ is typically the same for all spectral lines ($f_{\text{grad}} = 1.01 \pm 0.01$), which is consistent with no correction of the projection factor due to the velocity gradient. The uncertainty on $f_{\text{grad}}$ is derived from the errors on $a_0$ and $b_0$. 

Figure~\ref{triple}~(\emph{bottom left}) presents the interpoled $RV_{\text{c}}$ curve of $\beta\,$Cas for the FeI spectral line ($\lambda = 4\,508.288\,$\AA). We clearly see an increase of the amplitude of the radial velocity curve ($\sim 4.6 \pm 0.9 \, \%$ per cycle). Moreover, the radial velocity curves have several minima and maxima and we can easily deduce a period of pulsation. We find $P = 0.10046 \pm 0.00054\,\text{d}$. Our value agrees well with that of \citet{riboni_1994_P}.\\In Fig.~\ref{triple}~(bottom right), $\Delta RV_{\text{c}}$ is plotted as a function of $D$ for cycle 2 (see the vertical line in the figure). For $\beta\,$Cas, the range of the spectral line depth is eight times lower compared to AI Vel. The velocity gradient is $f_{\text{grad}} = 0.64 \pm 0.82$. The large error bar prevents any physical discussion about the exact value of $f_{\text{grad}}$. We note that $f_{\text{grad}}=0.64$ implies a huge velocity gradient in the star's atmosphere, which seems unrealistic. Since it is consistent with our determination, we assumed $f_{\text{grad}} =1.0$. Considering  $a_0 = 0$, we obtained $b_0 = 4.07 \pm 0.25$ with a reduced $\chi^2$ of 1.0, which provides an uncertainty on $f_{\text{grad}}$ of 0.14.
The $f_{\text{o-g}}$ correction, which is the last component of the projection factor decomposition, cannot be measured from observations. To estimate the differential velocity between the \emph{optical} and \emph{gas} layers at the photosphere of the star, we need a hydrodynamic model.\\ However, modelling the pulsating atmosphere of $\delta$ Scuti stars is not an easy task because of (1) cycle-to-cycle variations (non-radial modes) and (2) fast rotation in some cases (as for $\beta\,$Cas). However, $f_{\text{o-g}}$ have been studied intensively in the Cepheids \citep{nardetto_2004, fog_nardetto_2007, nardetto_2011}, and it seems that there is a linear relation between $f_{\text{o-g}}$ and $\log P$: $f_{\text{o-g}} = [-0.023 \pm 0.005]\,\log{P} + [0.979 \pm 0.005]$. Moreover, we have a theoretical value of $f_{\text{o-g}}$ for the short-period $\beta$-Cephei $\alpha\,$Lup ($P=0.2598\,$d), $f_{\text{o-g}} = 0.99 \pm 0.01$, which seems to be consistent with the $\log P$-$f_{\text{o-g}}$ relation of Cepheids \citep{alpha_lup_nardetto_2012}. For our study, we propose to extend this law for the $\delta\,$Scuti $\beta\,$Cas and AI Vel. Considering $P = 0.11157\,\text{d}$ for AI Vel and $P = 0.10046 \pm 0.00054\,\text{d}$ for $\beta\,$Cas (from this paper), we find $f_{\text{o-g}} = 1.00 \pm 0.02$ (0.02 is a conservative arbitrary uncertainty) for both stars, which basically means no photospheric correction to the projection factors.

\section{Discussion}\label{conclusion}

We can now calculate the projection factor $p$, using the relation $p = p_\mathrm{0} f_\mathrm{grad} f_{\text{o-g}}$. We find $p  = 1.44 \pm 0.05$ for AI Vel and $p = 1.41 \pm 0.25$ for $\beta\,$ Cas. However, the generalisation of this study to any $\delta\,$Scuti stars is presently limited since we have to study the impact of multi-modes, in particular non-radial ones, on the projection factor. This complicated effect has been studied by several authors for the bisector method of the radial velocity determination \citep{Dziembowski_1977, Balona_1979, Stamford_1981, Hatzes_1996}. We assume a star pulsating in two modes, one radial and one non-radial. The projection factor can be defined as $p = \alpha_{\mathrm{r}} p_{\mathrm{r}} + \alpha_{\mathrm{nr}} p_{\mathrm{nr}}$, with $\alpha_{\mathrm{r}}$ and $\alpha_{\mathrm{nr}}$ the relative contributions of the velocity amplitudes of the radial and non-radial modes to the pulsation ($\alpha_{\mathrm{r}} + \alpha_{\mathrm{nr}}=1$). $p_{\mathrm{r}}$ is our previous decomposition of the projection factor in the case of a purely radial mode $p_{\mathrm{r}} = p_\mathrm{0} f_\mathrm{grad} f_{\text{o-g}}$, while $p_{\mathrm{nr}} $ is the projection factor in the case of a purely non-radial mode. Using Eq. 4 of \cite{Hatzes_1996}, we find that $p_{\mathrm{nr}}= p_\mathrm{0} e ^{+k\ell^2}$, where $k=0.15$ in the case of a non radial p-mode and $k=1.2$ in the case of a non-radial g-mode. $\ell$ is the spherical harmonic degree. This relation is derived for the first moment (\emph{i.e.} the radial velocity) determination only, which is independent of the star's rotation. This equation can be used when $\ell=m$ only (where m is the spherical harmonic order). Additional work is necessary to derive it when  $\ell \neq m$. We emphasize also that the higher $\ell$($=m$), the higher is the non-radial projection factor. This is expected since for high values of $\ell$ (and this is qualitatively also true when $\ell \neq m$) there are more red- and blue-shifted velocity zones on the star that cancel each other in the integrated line profile, which leads to a lower amplitude of the non-radial velocity curve, and in turn a high value of the non-radial projection factor (see Eq. 1\&2 of \cite{Hatzes_1996}).

We performed our study on AI Vel and $\beta\,$Cas under the assumption of \textit{monoperiodic radial} pulsation. Our results (without non-radial correction) are consistent (at the 1$\sigma$ level) with the period-projection-factor ($Pp$) relation $p = [-0.071 \pm 0.020]\, \log P + [1.311 \pm 0.019]$ by \cite{laney_2009} applied for classical and dwarf Cepheids (it corresponds to $p = 1.38 \pm 0.02$ for AI Vel and $\beta\,$Cas).  To derive these values, Laney \& Joner  simply compared the distance of the stars obtained from the \emph{PL} relation with the distances from the photometric version of the Baade-Wesselink method. This suggests that the eventual non-radial components of $\beta\,$Cas have probably a negligeable impact on the projection factors (which means $\alpha_{\mathrm{nr}} \simeq 0)$. Interestingly, if we use the $Pp$ relation obtained for classical Cepheids by \cite{fog_nardetto_2007}, $ p = [-0.064 \pm 0.020]\, \log P + [1.376 \pm 0.023] $ (derived with the first moment method\footnote{We remind that using the cross-correlation method, one has to use the $Pp$ relation from \cite{nardetto_2009}.}) to derive the projection factors of the two $\delta$ Scuti stars, we find $p = 1.44 \pm 0.01$, which is consistent with our values. This seems to show that the $Pp$ relation provided by \cite{fog_nardetto_2007} for single lines is also applicable to $\delta\,$Scuti stars pulsating in a dominant \textit{radial} mode. In addition, and as already shown, for fast-rotating $\delta$ Scuti stars, an intrinsic dispersion of the $Pp$ relation due to the random orientation of the rotation axis has to be considered.
\begin{acknowledgements}
G.G. and N.N. thank J. Monnier for useful discussions. W.G. is grateful for support from the BASAL Centro de Astrofisica y Tecnologias Afines (CATA) PFP-06/2007. E.P. and M.R. acknowledge financial support from the Italian PRIN-INAF 2010 {\it Asteroseismology: looking inside the stars with space- and ground-based observations}.
\end{acknowledgements}

\bibliographystyle{aa}

\bibliography{sf2a-template}

\end{document}